\documentclass[letterpaper]{article} 
\usepackage{aaai2027} 
\nocopyright
\usepackage[hyphens]{url} 
\usepackage{graphicx} 
\usepackage{natbib} 
\usepackage{caption} 
\usepackage{booktabs}
\usepackage{makecell}
\usepackage{multirow}
\usepackage{amsmath}
\usepackage{amssymb}
\usepackage{siunitx}
\usepackage{cuted}
\usepackage{caption}

\newcommand{\best}[1]{\textbf{#1}}
\newcommand{\second}[1]{\underline{#1}}
\newcommand{\tablebodyfont}{\scriptsize}

\newcommand{\tablearraystretch}{1.02}
\newcommand{\method}{GazeFS}

\title{%
GazeFS: Target-Centered Gaze-Trajectory Forecasting and Stabilization from Gaze--Head History
}

\author{%
Yaozheng Xia (1), Zaiping Zhu (2), Bo Pang (3), Minghao Xie (4),\\
Hui Li (1), Shaorong Wang (1)\corresponding, Sheng Li (3)\corresponding%
}
\affiliations{%
(1) School of Information Science and Technology, School of Artificial Intelligence, Beijing Forestry University\\
(2) School of Digital Art, Nanjing University of the Arts\\
(3) School of Computer Science, Peking University\\
(4) Leicester International Institute, Dalian University of Technology%
}

\begin{document}
\maketitle
\begingroup
\setlength{\stripsep}{2pt plus 1pt minus 1pt}

\begin{strip}
  \centering

  \includegraphics[
    width=0.93\textwidth,
    keepaspectratio
  ]{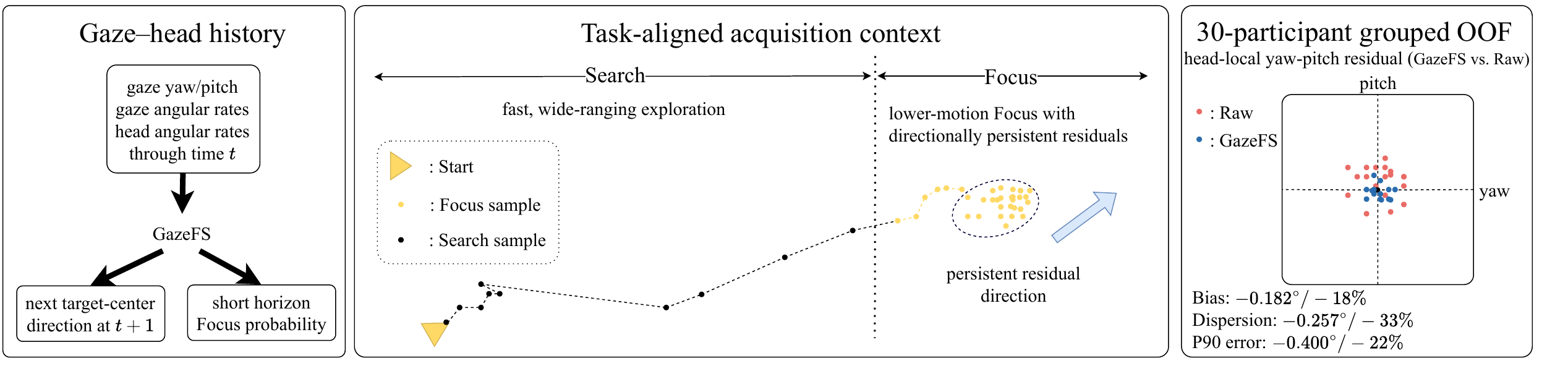}
  \par

  \parbox{0.95\textwidth}{%
    \captionof{figure}{
      Overview of \method. From task-aligned gaze--head history, the model
      forecasts the next target-center direction and short-horizon Focus
      probability without target geometry at inference; grouped OOF shows
      lower Focus bias, dispersion, and P90 error than Raw.
    }
    \label{fig:teaser}
  }
\end{strip}
\endgroup
\begin{abstract}
Target-centered gaze interaction requires more than suppressing frame-to-frame
fluctuations: target acquisition produces task-aligned changes in gaze-head
dynamics, while a gaze trace may retain a persistent target-relative residual
direction. We formulate gaze correction as online target-centered gaze-trajectory forecasting and stabilization
and introduce GazeFS, which maps a variable-length gaze-head history to the next
target-center direction and a short-horizon Search/Focus estimate without target
information at inference. Across 7,960 acquisition episodes from 30 participants,
Search--Focus differences remain stable under quality control, onset exclusion,
and duration matching. History windows improve phase
decoding over the current endpoint, but explicit task progress remains a strong
control. Under the 30-participant, five-fold grouped out-of-fold protocol across
three seeds, the reductions relative to raw hold in Focus episode bias,
within-episode dispersion, and P90 target error are $0.182^\circ$,
$0.257^\circ$, and $0.400^\circ$, with participant-bootstrap 95\% confidence
intervals excluding zero. Endpoint-free replay from empty history preserves the
Focus advantage and yields raw-network phase balanced accuracy/AUPRC of
0.925/0.993; coordinate controls further show that recent history contributes
beyond explicit progress metadata. GazeFS therefore improves Focus target
centering and empirical residual contraction while leaving temporal smoothness
as a separate objective.
\end{abstract}
\section{Introduction}
\label{sec:introduction}
Eye gaze enables fast, hands-free interaction in augmented and virtual
reality~\cite{plopski2022eye}. As a pointing signal, however, gaze estimates
reported by head-mounted devices exhibit two errors that should not be
conflated: short-timescale fluctuation, observed as local dispersion or
frame-to-frame motion, and persistent target-relative offset, which displaces
the interaction ray from the intended target
~\cite{aziz2022hololens,wagner2024eyehand,marquardt2024lighting}.
A low-pass filter can attenuate the former, but its steady-state output remains
centered on the measured signal and therefore cannot, by itself, infer a
corrected target center. A gaze trace can consequently be smooth yet systematically off target. Here,
``stabilization'' denotes target-relative centering and residual contraction; temporal smoothness is evaluated as a separate objective.

Target acquisition also evolves as a user locates a target, approaches it, and
maintains a target-centered direction for confirmation. We operationalize this
task progression as two confirmation-anchored phases, \emph{Search} and
\emph{Focus}; they are task-defined labels rather than assumed physiological
states. This raises a joint AI question: can a target-agnostic online model use
gaze-head history to infer both \emph{where} the target center lies and
\emph{when} the acquisition has entered the short-horizon Focus phase?

Prior methods address only isolated aspects of gaze-based target acquisition. Temporal filters reduce rapid
fluctuation but remain anchored to current measurement
~\cite{spakov2012filters,kumar2008gaze}, while eye-head pointing combines instantaneous
motion cues
~\cite{sidenmark2019eyehead,sidenmark2020bimodalgaze,
sidenmark2022weighted,hou2024gazeswitch}. Spatial correction methods either infer recalibration from interaction correspondences or rely on scene geometry and target candidates to adjust the landing position~\cite{hou2025onlineeye,wei2023targetselection}.
In parallel, intent and adaptive-dwell models estimate when activation should occur without correcting the underlying pointing direction
~\cite{davidjohn2021intent,isomoto2022dwell,narkar2024gazeintent}.
Consequently, spatial correction and acquisition timing are typically modeled separately, often with runtime access to interaction feedback or candidate targets.

We motivate a joint formulation using synchronized gaze and head-motion recordings collected during user target-acquisition episodes. Complete-episode analyses show
robust Search-Focus differences in gaze-speed tails, head motion, and local
dispersion, while target-relative Focus residuals retain a persistent
episode-level direction. Controlled history-window probes show that recent gaze-head
context is more informative than a single current observation.
However, simple task-progress cues explain much of the phase
predictability, and perturbing frame order causes little loss.
We therefore treat history as acquisition-history context rather
than claiming that an exact temporal ordering is necessary.

These observations motivate three modeling choices. First, directionally
persistent Focus residuals require forecasting a target-center direction rather
than smoothing measured gaze. Second, phase-dependent behavior motivates a
parallel phase output and phase-aligned coordinate learning. Third, progressively
growing and variable-length histories motivate masked history encoding and
episode prefix refinement. We formulate \emph{task-state-aligned online target-direction
forecasting} and introduce \method. Given gaze--head observations through time
$t$, \method\ predicts the next-frame target-center direction in head-local
yaw--pitch and whether the next short interval contains Focus. Fine- and
coarse-scale tokens share a Transformer encoder with a causal mask and feed parallel trajectory and
phase heads. The predicted phase does not gate the trajectory decoder at
inference; task-phase labels instead shape coordinate learning during training. A two-stage refinement module addresses the short context
available at buffer initialization. Target information is used for offline label construction,
supervision, sampling, and evaluation, but never enters the inference tensor.

We evaluate \method\ over all 30 participants using five participant-grouped
outer folds and three seeds. On common valid Focus support, Full \method\ reduces
episode bias, within-episode dispersion, median error, and P90 error relative
to raw hold by $0.182^\circ$, $0.257^\circ$, $0.109^\circ$, and
$0.400^\circ$, respectively; participant-level simultaneous intervals over
the four primary outcomes remain below zero. The same centering advantage
persists in endpoint-free replay from empty history. Additional controls show
that recent gaze--head history contributes spatial information beyond
observable progress cues, although the available probes do not isolate exact
frame order as the necessary mechanism. Future-suffix, exact-prefix, and
endpoint-free tests verify that the reported outputs do not depend on endpoint knowledge or unobserved future samples.

\paragraph{Contributions.}
Our contributions are:
\begin{itemize}

\item We formulate target-centered gaze--trajectory forecasting and stabilization, which uses gaze-head history to jointly estimate the target-center direction and stabilize the evolving gaze trajectory throughout target acquisition, without requiring target information at inference time.
\item We introduce \method, which combines multi-scale history encoding, parallel direction and phase prediction, and phase-aligned training for target-direction correction.
\item We establish an evaluation protocol, demonstrating improved Focus-stage target centering without access to future or endpoint information.
    
\end{itemize}

\section{Related Work}
\label{sec:related_work}

\paragraph{Gaze filtering and spatial correction.}
Temporal filters suppress sample-to-sample gaze fluctuation but generally
remain anchored to the measured direction
~\cite{plopski2022eye,spakov2012filters,kumar2008gaze}. Eye--head pointing
combines the rapid response of gaze with the relative stability of head motion
~\cite{sidenmark2019eyehead,sidenmark2020bimodalgaze,sidenmark2022weighted,hou2024gazeswitch}.
Candidate-aware methods instead infer an intended object from scene geometry or
learned endpoint distributions~\cite{pfeuffer2017gazepinch,wei2023targetselection}. Scene-level gaze-target detection further combines human-attention and activity
cues to localize the attended object~\cite{yang2024gazetarget}. These approaches improve smoothness, control, or object selection, but they
either remain tied to current measurements or require an explicit candidate
set. \method\ forecasts a target-centered direction from target-agnostic
gaze--head history. Target information is restricted to offline construction and
supervision, and endpoint-only and Endpoint+Progress controls test whether the
result can be explained by nonlinear current-frame or progress-conditioned
calibration.

\paragraph{Temporal gaze forecasting.}
Prior work uses gaze, head motion, and scene history to anticipate future
viewing behavior
~\cite{hu2019sgaze,burlingham2024dependencies,hu2020dgaze,
hu2021fixationnet,rolff2022gazetransformer}. Recent work further formulates
short-horizon VR gaze as multivariate time-series forecasting with
oculomotor-event prediction~\cite{melnyk2026timeseries}, while general
forecasting architectures model multi-scale or non-stationary temporal
patterns, including multi-scale
inter-series correlations~\cite{wu2023timesnet,cai2024msgnet}. Most prior forecasting methods predict subsequently observed
gaze for attention anticipation or latency compensation.
\method\ instead predicts a latent target-center interaction
direction whose geometry is available only during training. 

\paragraph{Fixation, intent, and task-aligned phase.}
Velocity- and dispersion-based methods identify physiological fixations and
saccades~\cite{salvucci2000identifying}, whereas intent and adaptive-dwell
models estimate interaction readiness or activation timing
~\cite{davidjohn2021intent,isomoto2022dwell,narkar2024gazeintent}. Our
confirmation-anchored Search/Focus labels encode target-acquisition progress,
not an intrinsic oculomotor regime. \method{} predicts this short-horizon phase
jointly with target direction from a shared historical representation. 

\section{Data and Task Definition}
\label{sec:dataset}
We analyze 7,960 60-Hz episodes from 30 HoloLens~2 participants with
\mbox{\(\mathbf{x}_t=[\mathbf g_t^\top,\dot{\mathbf g}_t^\top,
\dot{\mathbf h}_t^\top]^\top\in\mathbb{R}^7\)}.
\par\noindent
The channels are head-local gaze yaw--pitch, their angular rates, and head
yaw--pitch--roll rates.
The target-center direction is expressed in the same frame but is used only for
offline label construction, supervision, and evaluation; it is never provided
at inference.

\begin{figure}[htbp]
    \centering
    \includegraphics[width=0.99\linewidth]{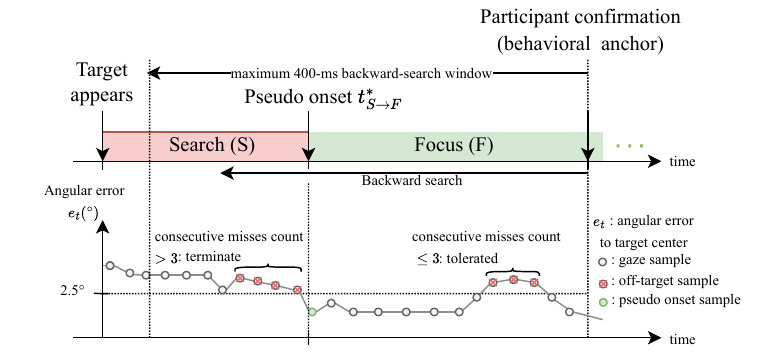}
    \caption{Confirmation-anchored construction of the Search
    and Focus task-phase labels. Starting from participant
    confirmation, a backward scan locates the earliest stable
    target-centered segment; its onset defines the Search-to-Focus
    boundary. The angular, continuity, and search-window
    thresholds are shown in the figure and detailed in
    Supplement Sec.~B.}
    \label{fig:task_label_construction}
\end{figure}

Figure~\ref{fig:task_label_construction} summarizes the label construction. We derive two task-aligned phase labels, \emph{Search} and
\emph{Focus}, from each confirmation-anchored gaze-to-target
trace. Treating participant confirmation as a retrospective
anchor, a backward scan locates the earliest stable
target-centered segment. Frames before this onset are labeled
Search and frames from the onset onward are labeled Focus.
These labels represent acquisition progress rather than
physiological fixation or saccade states. Collection, coordinate conversion, quality control, threshold sensitivity,
and participant-grouped splits are detailed in Supplement Sec. B.

The following analysis uses complete episodes to characterize phase-dependent
gaze--head structure. Natural history-window probes use frozen endpoints across
endpoint-only, ordered-history, order-invariant, and explicit-progress controls;
the complete protocol is reported in Supplement Sec. A.2.

\paragraph{Ethics and consent.}
Both studies were approved by the responsible institutional ethics committee
(identifier withheld for anonymous review), and all participants provided
informed consent before participation.

\section{Task-Aligned Gaze--Head Structure}
\label{sec:dataset_preparation}

\paragraph{Task-aligned non-stationarity.}
Table~\ref{tab:task_structure}(a) shows that Focus has a lower gaze-speed tail,
lower head motion, and lower local dispersion than Search. The effects retain
their direction after excluding $\pm300$\,ms around the recovered boundary and
after matching the Search and Focus duration distributions. Sensitivity of the
label construction and retained membership to threshold choices is reported
separately in Supplement Sec.~B.3. Together, these results support
task-aligned non-stationarity during acquisition.
\begin{table}[t]
\centering
\begingroup
\tablebodyfont
\renewcommand{\arraystretch}{\tablearraystretch}
\setlength{\tabcolsep}{2.0pt}
\textbf{(a) Focus $-$ Search effects.}\par\vspace{1pt}
\begin{tabular}{@{}l
S[table-format=-2.3]
S[table-format=-2.3]
S[table-format=-2.3]@{}}
\toprule
\textbf{Metric} & {\textbf{Default}} & {\textbf{Excl.}} & {\textbf{Match}} \\
\midrule
Gaze P90 ($^\circ$/s) & -15.263 & -14.705 & -17.872 \\
Head med. ($^\circ$/s) & -2.291 & -1.860 & -2.421 \\
Dispersion ($^\circ$) & -0.562 & -0.404 & -0.567 \\
\bottomrule
\end{tabular}

\vspace{2pt}
\textbf{(b) Raw Focus residual persistence.}\par\vspace{1pt}
\setlength{\tabcolsep}{2.4pt}
\begin{tabular}{@{}l
S[table-format=1.3]
l@{}}
\toprule
\textbf{Metric} & {\textbf{Estimate}} & \textbf{95\% CI} \\
\midrule
Early/late bias-vector cosine $\uparrow$ & 0.886 & [0.856, 0.912] \\
Persistent-offset ratio $\uparrow$ & 0.977 & [0.972, 0.982] \\
\bottomrule
\end{tabular}

\vspace{2pt}
\textbf{(c) Natural-window phase diagnostic.}\par\vspace{1pt}
\setlength{\tabcolsep}{2.1pt}
\begin{tabular}{@{}l
S[table-format=1.3]
S[table-format=1.3]
S[table-format=1.3]@{}}
\toprule
\textbf{Input} & {\textbf{BA$_{64}$ $\uparrow$}} & {\textbf{BA$_{128}$ $\uparrow$}} & {\textbf{AUPRC$_{128}$ $\uparrow$}} \\
\midrule
Endpoint only & 0.640 & 0.640 & 0.881 \\
Ordered history & 0.927 & 0.928 & 0.995 \\
Unordered summary & 0.914 & 0.921 & 0.993 \\
Elapsed time + endpoint & 0.928 & 0.928 & 0.997 \\
Elapsed time + ordered & 0.925 & 0.924 & 0.996 \\
\bottomrule
\end{tabular}
\caption{Task-aligned gaze--head structure. Panel (a) uses all 30
participants; ``Excl.'' removes $\pm300$\,ms around the Search/Focus boundary
and ``Match'' matches phase duration. Panel (b) summarizes direction
persistence within raw Focus over 7,960 episodes. Panel (c) is a one-seed
controlled probe; a three-seed 16/32-frame shuffle audit and paired effects
appear in Supplement Sec.~A.2.}
\label{tab:task_structure}
\endgroup
\end{table}

\paragraph{Directionally persistent Focus residuals.}
For each episode, we compute the mean raw target-relative residual vector
separately over the first and second halves of Focus. Table~\ref{tab:task_structure}(b)
shows an early/late bias-vector cosine of $0.886$ and a persistent-offset
ratio of $0.977$, with both intervals well above zero. The residual therefore
tends to retain a similar direction across Focus rather than behave as purely
frame-local fluctuation, motivating history-based target-center forecasting
rather than local smoothing alone (Supplement Sec.~A.1).

\paragraph{Recent history provides acquisition context.}
Table~\ref{tab:task_structure}(c) shows that a recent gaze--head window raises
balanced accuracy from $0.640$ for the current 7D observation to
$0.927/0.928$ at 64/128 frames and remains above a compact order-invariant
summary of the same window. However, historically observed elapsed-time features
are competitive with the full window, and preserving the same observations
while shuffling individual frames or short blocks produces essentially no
loss. These probes therefore show that access to rich recent-frame content is
useful, without identifying exact frame order as the necessary mechanism.
This still motivates variable-length history modeling because multiple
recent observations are more informative than a single endpoint; later
coordinate-level OOF controls test whether the spatial gains persist beyond
explicit progress cues. Exact probe definitions and paired effects appear in
Supplement Sec.~A.2.

\section{Online Target-Centered Gaze-Trajectory Forecasting and Stabilization}
\label{sec:method}

\subsection{Problem Formulation}
For sample $i$, let $\overline{\mathbf X}_i\in\mathbb R^{T\times 7}$
be a standardized, padded gaze--head tensor whose first
$\ell_i\leq T$ rows form the observed history and whose
remaining rows are padding. The seven inference channels are gaze yaw--pitch, their angular rates,
and head yaw--pitch--roll rates. Given only the observed prefix
$\overline{\mathbf X}_{i,1:\ell_i}$---that is, the currently
available history---\method\ predicts a
dense target-direction sequence and patch-level Search/Focus logits:
\begin{equation}
(\widehat{\mathbf A}_i,\mathbf L_i^{\mathrm{ph}})
=f_{\boldsymbol\theta}(\overline{\mathbf X}_i,\ell_i).
\label{eq:method_mapping}
\end{equation}
$\widehat{\mathbf A}_i$ is the target-center direction expressed in
head-local yaw--pitch. It is not a reconstruction of future recorded gaze.
No target-derived quantity enters $f_{\boldsymbol\theta}$ at inference;
target information and task labels are used only to construct supervision,
training losses, sampling or augmentation weights, and offline metrics.
The observed-history length $\ell_i$ and historical buffer-age/prefix
descriptors are also available to the model; although online
observable, they may encode task progress.

Indexing frames from zero, token $n$ is anchored at
$a_n=nS$ and predicts the next $S$ target-center directions,
aligned to frames $a_n+\delta$ for $\delta=1,\ldots,S$.
Its phase target asks whether the same short future interval
contains Focus. Thus the phase output is a short-horizon estimate,
not a current-frame label. The two outputs share a representation but remain
parallel; predicted phase never gates the direction forecast.

\subsection{\method\ Architecture}
\begin{figure*}[!htbp]
    \centering
    \includegraphics[width=0.95\textwidth]{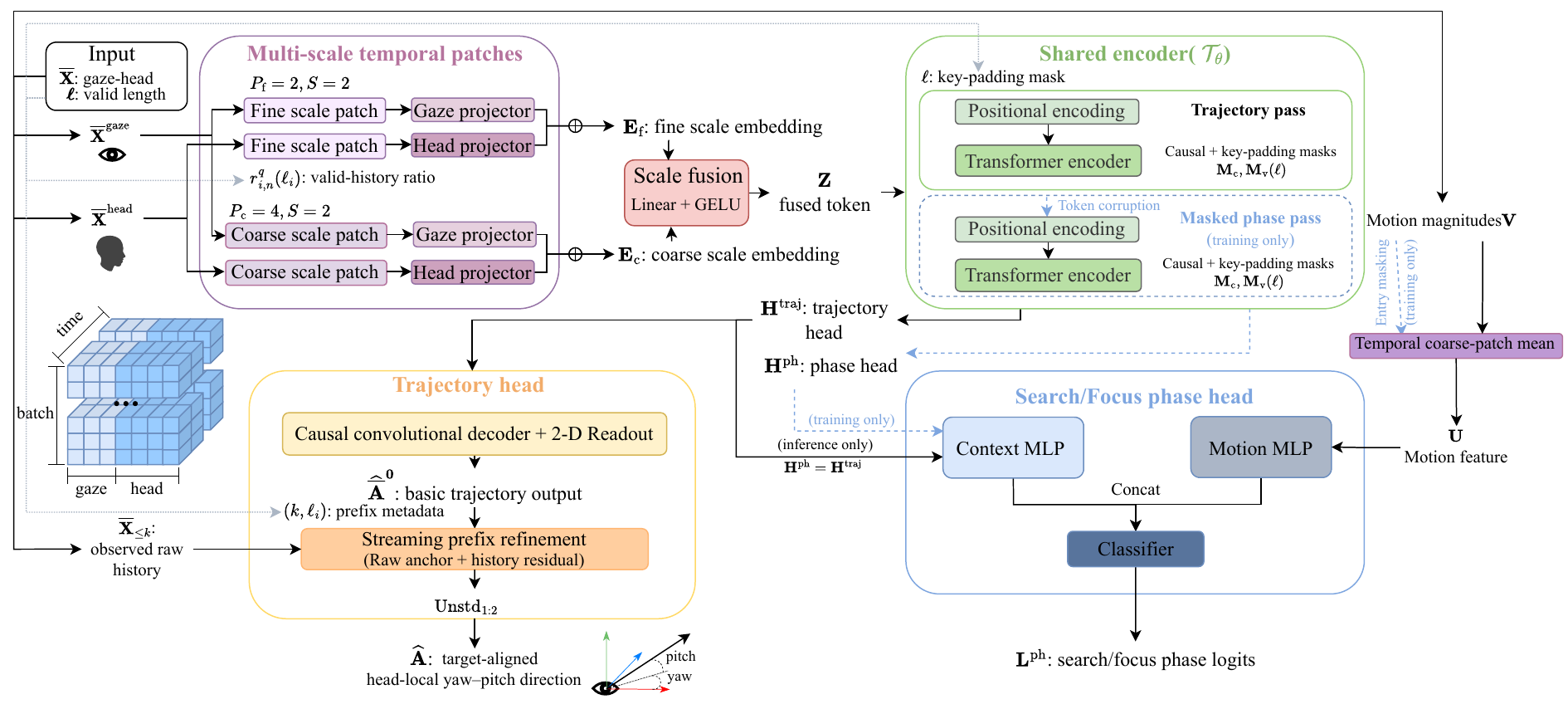}
    \caption{Overview of \method. Separate gaze and head projectors form
    fine- and coarse-scale temporal tokens, which are fused and processed by a
    shared Transformer encoder with a causal mask. The trajectory head forecasts target-aligned
    yaw--pitch and applies short-prefix refinement; the Search/Focus phase head
    combines the encoded context with temporally pooled motion magnitudes. The two heads are parallel: predicted phase is not fed to the
    trajectory branch. 
    Solid paths are used at inference, blue dashed paths are training-only,
    and dotted paths carry valid-history metadata. Target information and phase
    labels supervise training but never enter the inference graph.}
    \label{fig:method_overview}
\end{figure*}
Figure~\ref{fig:method_overview} shows the deployed path and training-only
regularization. The design addresses three constraints suggested by the data: gaze and head motion evolve at different local scales, inference must operate online without access to future observations, and only limited history is available immediately after initialization or reset.

\paragraph{Multi-scale temporal tokens.}
We split the input into a four-channel gaze stream and a
three-channel head-motion stream. Fine and coarse patches use
$P_f=2$ and $P_c=4$ with shared stride $S=2$, which provides a balanced development trade-off among coordinate error, phase quality, and output support (Supplement Sec.~D.2). At scale
$q\in\{f,c\}$, token $n$ is constructed from the $P_q$
observations ending at anchor $a_n$. Missing pre-sequence
context is filled by replicating the first observed frame, and
$r^q_{i,n}(\ell_i)$ records the fraction of patch positions
supported by actual observed history rather than replicated
left-boundary context. Separate gaze and head projections are
added within each scale; aligned fine and coarse tokens are
then concatenated and fused into tokens $\mathbf Z$. The fine scale captures
rapid acquisition changes, while the coarse scale provides longer local motion context.

\paragraph{Shared encoder and parallel heads.}
The shared Transformer~\cite{vaswani2017attention} uses sinusoidal positional
encoding, an upper-triangular causal mask $\mathbf M_{\mathrm c}$, and a
key-padding mask $\mathbf M_{\mathrm v}(\boldsymbol\ell)$. Here, "causal" refers only to left-looking operators that
exclude future samples. The causal mask prevents each token from attending to later
anchors, whereas the key-padding mask excludes token positions
outside the observed history, allowing variable-length samples
to share a mini-batch. Its clean
trajectory representation is
\begin{equation}
\mathbf H^{\mathrm{traj}}
=\mathcal T_{\theta}(\mathbf Z+\operatorname{PE};
\mathbf M_{\mathrm c},\mathbf M_{\mathrm v}(\boldsymbol\ell)).
\label{eq:htraj}
\end{equation}
During training, a random fixed-cardinality token mask $\mathbf M_p$ forms
\begin{equation}
\mathbf Z^{e}=(1-\mathbf M_p)\odot\mathbf Z
+\mathbf M_p\odot\mathbf e_{\mathrm{mask}},
\label{eq:corrupted_tokens}
\end{equation}
and the same encoder parameters produce
\begin{equation}
\mathbf H^{\mathrm{ph}}=
\begin{cases}
\mathcal T_{\theta}(\mathbf Z^{e}+\operatorname{PE};
\mathbf M_{\mathrm c},\mathbf M_{\mathrm v}(\boldsymbol\ell)),
& \text{training},\\
\mathbf H^{\mathrm{traj}}, & \text{inference}.
\end{cases}
\label{eq:hph}
\end{equation}
Thus deployment uses one encoder pass. The trajectory head always consumes
$\mathbf H^{\mathrm{traj}}$. During training, the Search/Focus
phase head consumes $\mathbf H^{\mathrm{ph}}$ from the
masked second pass; at inference it reuses
$\mathbf H^{\mathrm{traj}}$. The phase logits are never fed to
the trajectory head, so deployment requires only one encoder
pass.

\paragraph{Trajectory head and streaming-prefix refinement.}
A causal convolutional decoder maps $\mathbf H^{\mathrm{traj}}$ to a
standardized base forecast
$\widehat{\overline{\mathbf A}}^{0}\in
\mathbb R^{B\times N\times S\times2}$. Concatenating successive token outputs
gives the dense alignment
$\widehat{\mathbf a}_{i,k}\leftrightarrow\mathbf a^*_{i,k+1}$; for clarity, the notation retains only the two yaw--pitch channels used for
supervision.

For the first $R$ dense outputs, two causal residual modules refine the base
forecast in the standardized domain:
\begin{align}
\overline{\mathbf a}^{(1)}_{i,k}
&=(1-\rho_k)\widehat{\overline{\mathbf a}}^{0}_{i,k}
 +\rho_k\bigl(\overline{\mathbf x}_{i,k,1:2}
 +\boldsymbol\Delta^r_{i,k}\bigr),\\
\widehat{\overline{\mathbf a}}_{i,k}
&=\overline{\mathbf a}^{(1)}_{i,k}
 +\gamma_{i,k}\boldsymbol\Delta^h_{i,k}.
\label{eq:prefix_refinement}
\end{align}
The raw-anchor residual $\boldsymbol\Delta^r$ uses raw observed gaze--head history. The
history residual $\boldsymbol\Delta^h$ additionally receives the first-stage
direction and prefix descriptors derived from $k$ and $\ell_i$. Gates
$\rho_k$ and $\gamma_{i,k}$ decay as context accumulates, leaving the base
forecast unchanged outside the short prefix. The final yaw--pitch output is
obtained by inverse standardization.

\paragraph{Motion-aware Search/Focus phase head.}
From standardized rates, we form gaze- and head-motion magnitudes and temporally
mean-pool them over the coarse patches, yielding
$\mathbf U\in\mathbb R^{B\times N\times2}$. A context MLP consumes
$\mathbf H^{\mathrm{ph}}$, a motion MLP consumes $\mathbf U$, and their
features are concatenated:
\begin{equation}
\mathbf L^{\mathrm{ph}}_{i,n,:}
=D_{\mathrm{ph}}\!\left(
[\,F_t(\mathbf H^{\mathrm{ph}}_{i,n});F_v(\mathbf U_{i,n})\,]
\right).
\label{eq:phase_head}
\end{equation}
Training-only token replacement and motion-entry masking discourage reliance
on isolated token and rate measurements; inference reuses
$\mathbf H^{\mathrm{traj}}$ and the unmasked motion descriptors.

\subsection{Task-Aligned Learning}

Let $y_{i,t}\in\{0,1\}$ denote the frame-level Search/Focus label. The native phase target for token $n$ is
the next-$S$ maximum
\begin{equation}
\bar y_{i,n}=\max_{1\leq j\leq S}y_{i,a_n+j}.
\label{eq:next_s_state_target}
\end{equation}
For frame-level evaluation, each token probability is expanded over its
corresponding $S$ outputs and aligned to the physical $t+1$ frame label; the
native training target remains the next-$S$ maximum above.
The coordinate target is the next-frame target-center direction
$\mathbf a^*_{i,k+1}$. Let $\widetilde y_{i,k}\in[0,1]$ be a label-derived
Focus weight softened only near the task boundary, and define
\begin{equation}
\mathbf e_{i,k}=
[\operatorname{wrap}(\widehat\psi_{i,k}-\psi^*_{i,k+1}),
\widehat\phi_{i,k}-\phi^*_{i,k+1}]^\top.
\label{eq:coordinate_residual_compact}
\end{equation}
The coordinate objective combines a phase-weighted Huber term for robust,
component-wise yaw/pitch displacement with a directional cosine term for
scale-invariant orientation mismatch in the yaw--pitch plane,
\begin{equation}
\mathcal L_{\mathrm{coord}}=
\left\langle w_r(\widetilde y)
\left[w_H(\widetilde y)H_{\delta(\widetilde y)}(\mathbf e)
+w_C(\widetilde y)\ell_{\cos}\right]\right\rangle_v,
\label{eq:coordinate_objective_compact}
\end{equation}
where $\langle\cdot\rangle_v$ averages valid outputs. Crucially,
$\widetilde y$ comes from $y$, never from the predicted phase. 

The phase head
uses $\alpha$-balanced focal-weighted cross-entropy~\cite{lin2017focal}. Two
training-only geometry terms penalize missed target-plane entry on raw-stable
Focus frames and around synthetic boundaries. The complete objective is
\begin{equation}
\begin{split}
\mathcal L={}&e^{-s_c}\mathcal L_{\mathrm{coord}}+s_c
+e^{-s_f}\lambda_f\mathcal L_{\mathrm{focus}}+s_f\\
&+\lambda_e\mathcal L_{\mathrm{effective}}
+\lambda_b\mathcal L_{\mathrm{boundary}},
\end{split}
\label{eq:joint_objective}
\end{equation}
with homoscedastic task weighting~\cite{kendall2018multitask}. Exact masks,
augmentation, geometry construction, tensor alignment, coefficients, and the
export/runtime profile are given in Supplement Sec. C.

\section{Experiments}
\label{sec:experiments}

\subsection{Protocols, Statistics, and Baselines}
\paragraph{Evaluation and grouped-OOF protocol.}
Fixed-window evaluation uses common 128-frame crops; exact-prefix evaluation
scores the next frame after $L$ observed frames; and endpoint-free replay starts
from empty history, grows to 128 frames, and then slides without endpoint or
future information. Five participant-grouped outer folds use 21/3/6
participants for fitting, checkpoint selection, and testing. Scaling, training,
and selection exclude outer-test participants; learned methods use seeds 42,
314, and 2718. The fixed 24/3/3 split and consecutive-target reset tests are
supplementary development/lifecycle diagnostics.

\paragraph{Inference contract and controls.}
All methods estimate direction at $t+1$ from information available through
$t$. Raw-hold copies the measured direction at $t$; One
Euro~\cite{casiez2012oneeuro} filters the same stream with frozen parameters.
Endpoint-only uses $\mathbf{x}_t$, whereas Endpoint+Progress adds observable
buffer-age/history-ratio descriptors but no earlier gaze--head samples. The
no-explicit-progress variant retains history while removing explicit
history-ratio, positional, and prefix metadata. The Transformer-Causal baseline uses the same participant folds, 7D inputs,
next-frame coordinate target, phase-label alignment, feature scaling, and
documented shared sampling and loss settings. Encoder-specific modules differ
by design; the audited matching scope is reported in Supplement Sec.~A.9. OOF results
use raw network outputs without EMA, dwell, or FSM decisions.

\paragraph{Metrics and participant-level inference.}
Fixed-window contrasts use the same finite, label-valid Focus samples for all
compared methods. Episode bias is the norm of the mean signed target-relative
residual, dispersion is its centered spatial spread, and target-error median
and P90 summarize typical and upper-tail angular distance. Temporal behavior is
reported separately as frame-to-frame residual motion, the mean
tangent-plane displacement between physically adjacent, jointly valid source
frames; it is not target error, dispersion, input velocity, or deployed pointer
jitter. Metrics are computed
per episode, summarized within participant and seed, averaged across the three
seeds within participant, and then averaged with equal participant weight.
Paired effects use 5,000 participant-bootstrap resamples. The four
Full-minus-Raw Focus spatial outcomes form the primary family, with
max-$|T|$ simultaneous intervals. Phase metrics use the physical $t+1$ label,
raw Focus probabilities, and threshold 0.5. Exact formulas and support rules
are in Supplement Sec.~A.3.

\subsection{RQ1: Focus-Stage Target Centering}

\begin{table}[t]
\centering
\begingroup
\tablebodyfont
\setlength{\tabcolsep}{2.2pt}
\renewcommand{\arraystretch}{\tablearraystretch}
\begin{tabular}{@{}l
S[table-format=1.3]
S[table-format=1.3]
S[table-format=1.3]
S[table-format=1.3]@{}}
\toprule
\textbf{Method} & {\textbf{Bias $\downarrow$}} & {\textbf{Disp. $\downarrow$}} &
{\textbf{Med. $\downarrow$}} & {\textbf{P90 $\downarrow$}} \\
\midrule
Raw hold & 1.010 & 0.777 & 1.016 & 1.814 \\
One Euro & 1.004 & 0.761 & 1.008 & 1.802 \\
Endpoint-only & 1.045 & 0.820 & 1.070 & 1.907 \\
Transformer-Causal & 0.968 & \second{0.570} & 1.048 & \second{1.599} \\
\method\ w/o align. & \second{0.908} & 0.642 & \second{1.003} & 1.626 \\
\textbf{Full \method} & \best{0.828} & \best{0.520} &
\best{0.906} & \best{1.415} \\
\bottomrule
\end{tabular}
\caption{Common-support fixed-window Focus outcomes (degrees) under grouped
OOF. Bold/underline denote best/second-best values; paired and simultaneous
intervals are in Supplement Sec.~A.4.}
\label{tab:oof_spatial}
\endgroup
\end{table}

On common Focus support, Full \method\ reduces Raw bias, dispersion, median
error, and P90 error by $0.182^\circ$, $0.257^\circ$, $0.109^\circ$, and
$0.400^\circ$, respectively; participant-bootstrap and simultaneous intervals
remain below zero. All four outcomes also improve relative to One Euro,
endpoint-only, Transformer-Causal, and \method\ without phase alignment. These
controls test current-frame calibration, a generic history model, and
phase-dependent coordinate training, respectively.

The gain is not temporal smoothing. Full \method\ increases mean
frame-to-frame target-relative residual motion by $0.0578^\circ$ per adjacent
physical frame and residual path by $7.003^\circ$ relative to Raw
(Supplement Sec.~A.4). These metrics quantify temporal movement of the residual
trajectory rather than spatial spread or device-level pointer jitter.

\subsection{RQ2: Phase Dependence and Mechanism Boundaries}

\begin{table}[t]
\centering
\begingroup
\tablebodyfont
\setlength{\tabcolsep}{1.8pt}
\renewcommand{\arraystretch}{\tablearraystretch}
\begin{tabular}{@{}lcccc@{}}
\toprule
\textbf{Phase} & \textbf{Bias $\downarrow$} & \textbf{Disp. $\downarrow$} & \textbf{Med. $\downarrow$} &
\textbf{P90 $\downarrow$} \\
\midrule
Search & 13.887/\best{13.475} & \best{1.687}/2.420 &
14.407/\best{14.064} & \best{16.306}/16.443 \\
Transition & 3.229/\best{3.057} & \best{4.192}/4.225 &
\best{2.113}/2.317 & \best{8.564}/8.572 \\
Focus & 1.010/\best{0.828} & 0.777/\best{0.520} &
1.016/\best{0.906} & 1.814/\best{1.415} \\
\bottomrule
\end{tabular}
\caption{Phase harm audit, shown as Raw/Full \method\ in degrees. Bold marks
the lower value within each pair; only Focus improves jointly across all four
spatial outcomes. Complete motion diagnostics appear in Supplement Sec.~A.4.}
\label{tab:phase_harm_audit}
\endgroup
\end{table}

Table~\ref{tab:phase_harm_audit} localizes the joint spatial improvement to
Focus; Search and transition do not improve consistently across centering,
spread, tail error, and residual motion. Because the no-alignment control keeps
the phase head but removes phase-dependent coordinate weighting, it supports
the coordinate-training mechanism rather than an isolated classifier benefit.
Development-only modality ablations show that gaze orientation carries target
bearing, while gaze/head rates mainly improve endpoint-free upper-tail error
(Supplement Sec.~D).

\subsection{RQ3: Online Phase, Progress Controls, and Replay}

\begin{table}[t]
\centering
\begingroup
\tablebodyfont
\setlength{\tabcolsep}{1.05pt}
\renewcommand{\arraystretch}{\tablearraystretch}
\begin{tabular}{@{}l
S[table-format=1.3]
S[table-format=1.3]
S[table-format=1.3]
S[table-format=1.3]
c
S[table-format=2.2]@{}}
\toprule
\textbf{Method} & {\textbf{BA $\uparrow$}} & {\textbf{F1 $\uparrow$}} & {\textbf{AUPRC $\uparrow$}} &
{\textbf{Cov. $\uparrow$}} & \makecell{\textbf{Onset}\\\textbf{Median/P90 $\downarrow$}} &
{\makecell{\textbf{Pre-onset}\\\textbf{Focus (\%)}}} \\
\midrule
Endpoint-only & 0.530 & 0.172 & 0.875 & 0.399 & 398/1368 & 3.79 \\
Transformer-Causal & 0.896 & \best{0.940} & 0.989 & 1.000 & \best{225}/630 & 13.93 \\
\method\ w/o align. & \second{0.901} & 0.934 & 0.990 & 1.000 &
\second{246}/\second{594} & 10.83 \\
\textbf{Full \method} & \best{0.903} & \second{0.936} & 0.990 & 1.000 &
248/\best{585} & 10.58 \\
\bottomrule
\end{tabular}
\caption{Raw-logit phase/onset outcomes. Cov. denotes threshold-crossing
coverage; onset entries are absolute median/P90 errors (ms). Pre-Focus is a
one-sided, unranked diagnostic. No EMA, dwell, or FSM is applied.}
\label{tab:oof_phase}
\endgroup
\end{table}

Full obtains 0.903 balanced accuracy, 0.936 Focus F1, 0.990 AUPRC, and
248/585\,ms median/P90 absolute onset error from raw phase outputs. This is a
trade-off rather than uniform dominance, and coverage is not a false-alarm
estimate because every source episode eventually acquires a target.

\begin{table}[t]
\centering
\begingroup
\tablebodyfont
\setlength{\tabcolsep}{1.7pt}
\renewcommand{\arraystretch}{\tablearraystretch}
\begin{tabular}{@{}lrrS[table-format=2.3]l@{}}
\toprule
\makecell{\textbf{$L$}\\\textbf{frames / s}} & \textbf{Valid} & \textbf{Cold} &
{\makecell{\textbf{Full}\\\textbf{median $\downarrow$}}} &
\makecell[l]{\textbf{$\Delta$ med. [95\% CI]}\\\textbf{(Full $-$ Raw)}} \\
\midrule
8 / 0.13   & 7,959 & 5,445 & 14.036 & $-0.154$ [$-0.180,-0.129$] \\
16 / 0.27  & 7,959 & 5,445 & 12.443 & $-0.169$ [$-0.203,-0.135$] \\
32 / 0.53  & 7,960 & 5,445 & 1.742 & $0.000$ [$-0.044,0.043$] \\
64 / 1.07  & 7,960 & 5,445 & 0.970 & $-0.086$ [$-0.137,-0.035$] \\
128 / 2.13 & 7,828 & 7,828 & 0.928 & $-0.142$ [$-0.202,-0.081$] \\
\bottomrule
\end{tabular}
\caption{Exact source-prefix evaluation in degrees at approximately 60 Hz.
Prefix $L$ scores frame $L$ from observations through $L-1$ and never counts
backward from a known endpoint. ``Cold'' excludes previous-target context.}
\label{tab:oof_prefix}
\endgroup
\end{table}

All 60 learned OOF checkpoints satisfy future-suffix invariance:
replacing samples strictly after an audit frame leaves every earlier
direction output and phase logit unchanged, with a global maximum
absolute difference of zero (Supplement Sec.~A.3). Exact-prefix
results improve over Raw at 8, 16, 64, and 128 frames, while the 32-frame
interval crosses zero; cold-start superiority is therefore not uniform.

\begin{table}[t]
\centering
\begingroup
\tablebodyfont
\setlength{\tabcolsep}{1.9pt}
\renewcommand{\arraystretch}{\tablearraystretch}

\textbf{(a) Endpoint-free Focus outcomes.}\par\vspace{1pt}
\begin{tabular}{@{}l
S[table-format=1.3]
S[table-format=1.3]
S[table-format=1.3]
S[table-format=1.3]@{}}
\toprule
\textbf{Method} & {\textbf{Bias $\downarrow$}} & {\textbf{Disp. $\downarrow$}} & {\textbf{Med. $\downarrow$}} & {\textbf{P90 $\downarrow$}} \\
\midrule
Raw hold & 1.010 & 0.777 & 1.016 & 1.814 \\
Transformer-Causal & \second{0.964} & \second{0.558} & \second{1.039} & \second{1.580} \\
\textbf{Full \method} & \best{0.824} & \best{0.508} & \best{0.900} & \best{1.397} \\
\bottomrule
\end{tabular}

\vspace{3pt}
\textbf{(b) Raw-network phase output.}\par\vspace{1pt}
\begin{tabular}{@{}l
S[table-format=1.3]
S[table-format=1.3]
S[table-format=2.1]@{}}
\toprule
\textbf{Method} & {\makecell{\textbf{Balanced}\\\textbf{accuracy $\uparrow$}}} &
{\textbf{AUPRC $\uparrow$}} &
{\makecell{\textbf{Pre-onset}\\\textbf{Focus (\%)}}} \\
\midrule
Transformer-Causal & \second{0.920} & \second{0.992} & 9.0 \\
\textbf{Full \method} & \best{0.925} & \best{0.993} & 6.0 \\
\bottomrule
\end{tabular}

\caption{Endpoint-free OOF from empty history. Context grows to 128 frames
and then slides without endpoint/future input, output EMA, or FSM decisions.
Pre-onset Focus is unranked; complete results appear in Supplement Sec.~A.6.}
\label{tab:endpoint_free_oof_main}
\endgroup
\end{table}

Table~\ref{tab:endpoint_free_oof_main} shows that the fixed-window centering
advantage persists when every episode is replayed from empty history. Relative
to Transformer-Causal, Full has higher balanced accuracy and AUPRC and a lower
onset-error P90, while Focus F1 is slightly lower and the median threshold
crossing is later; complete trade-offs are in Supplement Sec.~A.6.

\begin{table}[t]
\centering
\begingroup
\tablebodyfont
\setlength{\tabcolsep}{1.8pt}
\renewcommand{\arraystretch}{\tablearraystretch}
\begin{tabular}{@{}lcc
S[table-format=1.3]
S[table-format=1.3]
S[table-format=1.3]
S[table-format=1.3]@{}}
\toprule
\textbf{Method} &
\makecell{\textbf{Earlier}\\\textbf{history}} &
\makecell{\textbf{Explicit}\\\textbf{progress}} &
{\textbf{Bias $\downarrow$}} &
{\textbf{Disp. $\downarrow$}} &
{\textbf{Med. $\downarrow$}} &
{\textbf{P90 $\downarrow$}} \\
\midrule
Endpoint-only & -- & -- & 1.045 & 0.820 & 1.070 & 1.907 \\
Endpoint+Progress & -- & \checkmark & 0.962 & 0.775 & 0.985 & 1.783 \\
\makecell[l]{\method\ w/o \\ \quad explicit progress metadata}
    & \checkmark & -- & \best{0.803} & \second{0.643} & \best{0.887} & \second{1.481} \\
\textbf{Full \method}
    & \checkmark & \checkmark & \second{0.824} & \best{0.508} & \second{0.900} & \best{1.397} \\
\bottomrule
\end{tabular}
\caption{Endpoint-free history--progress controls. History denotes preceding
7D observations; progress denotes buffer-age, history-ratio, positional, and
prefix metadata. The no-explicit variant still exposes structural prefix
support; paired intervals appear in Supplement Sec.~A.7.}
\label{tab:history_progress_main}
\endgroup
\end{table}

\paragraph{History and progress.}
Table~\ref{tab:history_progress_main} separates current observation, explicit
progress, and earlier gaze--head history under the same endpoint-free protocol.
Progress alone improves all four spatial outcomes over Endpoint-only. A history
model without explicit progress metadata then improves bias, dispersion,
median, and P90 over Endpoint+Progress by $0.159^\circ$, $0.132^\circ$,
$0.098^\circ$, and $0.303^\circ$, respectively, with participant-paired
intervals below zero. Restoring the full progress/position context primarily
reduces dispersion and P90 and improves frame-to-frame residual motion and phase/onset
reliability. The control is bundled and does not isolate exact frame order.

\paragraph{Mature history and temporal localization.}
At $\geq128$ frames, after buffer length has saturated and short-prefix
refinement has ended, Full remains better than Endpoint+Progress by
$0.139^\circ$, $0.100^\circ$, $0.076^\circ$, and $0.284^\circ$ in bias,
dispersion, median, and P90, respectively; all four paired intervals remain
below zero, while residual-motion and path intervals include zero. Relative to
Raw, P90 is worse during the first 250\,ms after Focus onset
($+0.360^\circ$) but better after 500\,ms ($-0.455^\circ$), localizing the
claim to sustained Focus rather than initial target discovery
(Supplement Sec.~A.8).

\subsection{Secondary Closed-Loop Feasibility}
\label{sec:live_user_study}

A 12-participant within-subject HoloLens~2 study compares the native gaze
pointer, One Euro, a Weighted Pointing implementation of Weighted
Pointer~\cite{sidenmark2022weighted}, and \method\ Residual over 384 scored
trials per condition. Completion is near ceiling. In the all-trial
$\min(T,5\,\mathrm{s})$ sensitivity, \method\ is $0.196$\,s faster than One
Euro (95\% CI $[-0.263,-0.124]$, $p_{\mathrm{Holm}}=0.0044$); comparisons with
the native pointer and Weighted Pointing cross zero. Because only \method\
resets history/state after selection or timeout, this compares complete
deployed conditions rather than isolating learned correction or reset
(Supplement Sec.~F).

\begin{table}[t]
\centering
\begingroup
\tablebodyfont
\setlength{\tabcolsep}{1.45pt}
\renewcommand{\arraystretch}{\tablearraystretch}
\begin{tabular}{@{}l
S[table-format=3.1]
S[table-format=1.3]
S[table-format=1.3]
S[table-format=1.3]@{}}
\toprule
\textbf{Condition} &
{\makecell{\textbf{Completion}\\\textbf{(\%) $\uparrow$}}} &
{\makecell{\textbf{Completed-target}\\\textbf{time (s) $\downarrow$}}} &
{\makecell{\textbf{Focus}\\\textbf{error (cm) $\downarrow$}}} &
{\makecell{\textbf{Focus}\\\textbf{dispersion (cm) $\downarrow$}}} \\
\midrule
HoloLens~2 native & 99.2 & 1.562 & 2.069 & 2.230 \\
One Euro & \second{99.7} & 1.732 & 2.082 & 2.317 \\
Weighted Pointing & 99.5 & \second{1.552} & \best{2.018} & \second{2.221} \\
\textbf{\method\ Residual} & \best{100.0} & \best{1.544} & \second{2.067} & \best{2.214} \\
\bottomrule
\end{tabular}
\caption{Twelve-participant study with 384 scored trials per condition.
Bold/underline are descriptive only. Completed-target time is conditioned on
completion; the all-trial sensitivity is in Supplement Sec.~F.}
\label{tab:live_user_study}
\endgroup
\end{table}

\paragraph{Discussion.}
The controls suggest complementary roles for history and progress. Recent
history contributes spatial information beyond explicit buffer-age descriptors,
and this advantage remains after the buffer reaches 128 frames and short-prefix
refinement has ended. Progress/position context chiefly reduces dispersion,
upper-tail error, and target-relative temporal motion while improving
phase/onset reliability. Near-null shuffle effects show that rich recent
observations matter, but do not isolate exact frame order as necessary. The
no-alignment comparison supports phase-dependent coordinate training, not an
independent benefit of the auxiliary classifier; predicted phase is never fed
to the trajectory output. Stable Raw-relative gains occur in sustained Focus
rather than initial target discovery. The live study supports feasibility and
a narrow One-Euro duration contrast, while the remaining Raw-relative motion
cost motivates explicit centering--smoothness optimization.

\section{Limitations}

Search and Focus are confirmation-anchored task phases, and target information is
used for supervision although it is absent at inference. The evidence supports
complementary history and progress context, not a strict order law. Evaluation covers segmented successful acquisitions from one HoloLens~2
task. A single-seed held-quadrant screen supports extrapolation within the
recorded geometry, while open-stream false activation and cross-layout,
session, task, and device generalization remain to be evaluated.

Because model development used a fixed split from the same corpus, grouped OOF
is internal validation rather than independent-cohort confirmation. The
closed-loop comparison does not isolate reset from learned correction, and
centering retains a Raw-relative motion cost. PC-side ONNX latency excludes
sensing, network transport, and Unity rendering (Supplement Sec.~C.8).
\section{Conclusion}
\method\ forecasts the next target-center direction and a short-horizon
Search/Focus phase from gaze--head history without target geometry at
inference. Participant-grouped OOF and endpoint-free replay show
sustained-Focus reductions in bias, dispersion, median, and upper-tail error,
while controls indicate complementary roles for recent history and observable
progress. A secondary closed-loop study supports feasibility and a narrow
duration benefit over One Euro; temporal smoothness and broader
continuous-stream generalization remain open.

\bibliography{references}

@article{plopski2022eye,
  title={The Eye in Extended Reality: A Survey on Gaze Interaction and Eye Tracking in Head-Worn Extended Reality},
  author={Plopski, Alexander and Hirzle, Teresa and Norouzi, Nahal and Qian, Long and Bruder, Gerd and Langlotz, Tobias},
  journal={ACM Computing Surveys},
  volume={55},
  number={3},
  articleno={53},
  pages={53:1--53:39},
  year={2022},
  doi={10.1145/3491207}
}

@inproceedings{aziz2022hololens,
  title={An Assessment of the Eye Tracking Signal Quality Captured in the {HoloLens 2}},
  author={Aziz, Samantha D. and Komogortsev, Oleg V.},
  booktitle={Proceedings of the 2022 Symposium on Eye Tracking Research and Applications},
  articleno={5},
  pages={5:1--5:6},
  year={2022},
  doi={10.1145/3517031.3529626}
}

@inproceedings{wagner2024eyehand,
  title={Eye-Hand Movement of Objects in Near Space Extended Reality},
  author={Wagner, Uta and Jacobsen, Andreas Asferg and Feuchtner, Tiare and Gellersen, Hans and Pfeuffer, Ken},
  booktitle={Proceedings of the 37th Annual ACM Symposium on User Interface Software and Technology},
  articleno={84},
  pages={84:1--84:13},
  year={2024},
  doi={10.1145/3654777.3676446}
}

@inproceedings{spakov2012filters,
  title={Comparison of Eye Movement Filters Used in {HCI}},
  author={{\v{S}}pakov, Oleg},
  booktitle={Proceedings of the Symposium on Eye Tracking Research and Applications},
  pages={281--284},
  year={2012},
  doi={10.1145/2168556.2168616}
}

@inproceedings{marquardt2024lighting,
  title={Selection Performance and Reliability of Eye and Head Gaze Tracking under Varying Light Conditions},
  author={Marquardt, Alexander and Steininger, Melissa and Trepkowski, Christina and Weier, Martin and Kruijff, Ernst},
  booktitle={2024 IEEE Conference Virtual Reality and 3D User Interfaces (VR)},
  pages={546--556},
  year={2024},
  organization={IEEE},
  doi={10.1109/VR58804.2024.00075}
}

@inproceedings{sidenmark2019eyehead,
  title={Eye\&Head: Synergetic Eye and Head Movement for Gaze Pointing and Selection},
  author={Sidenmark, Ludwig and Gellersen, Hans},
  booktitle={Proceedings of the 32nd Annual ACM Symposium on User Interface Software and Technology},
  pages={1161--1174},
  year={2019},
  doi={10.1145/3332165.3347921}
}

@inproceedings{sidenmark2020bimodalgaze,
  title={BimodalGaze: Seamlessly Refined Pointing with Gaze and Filtered Gestural Head Movement},
  author={Sidenmark, Ludwig and Mardanbegi, Diako and Ramirez Gomez, Argenis and Clarke, Christopher and Gellersen, Hans},
  booktitle={ACM Symposium on Eye Tracking Research and Applications},
  articleno={8},
  pages={8:1--8:9},
  year={2020},
  doi={10.1145/3379155.3391312}
}

@article{sidenmark2022weighted,
  title={Weighted Pointer: Error-Aware Gaze-Based Interaction through Fallback Modalities},
  author={Sidenmark, Ludwig and Parent, Mark and Wu, Chi-Hao and Chan, Joannes and Glueck, Michael and Wigdor, Daniel and Grossman, Tovi and Giordano, Marcello},
  journal={IEEE Transactions on Visualization and Computer Graphics},
  volume={28},
  number={11},
  pages={3585--3595},
  year={2022},
  doi={10.1109/TVCG.2022.3203096}
}

@article{hou2024gazeswitch,
  title={GazeSwitch: Automatic Eye-Head Mode Switching for Optimised Hands-Free Pointing},
  author={Hou, Baosheng James and Newn, Joshua and Sidenmark, Ludwig and Khan, Anam Ahmad and Gellersen, Hans},
  journal={Proceedings of the ACM on Human-Computer Interaction},
  volume={8},
  number={ETRA},
  articleno={227},
  pages={227:1--227:20},
  year={2024},
  doi={10.1145/3655601}
}

@inproceedings{hou2025onlineeye,
  title={{Online-EYE}: Multimodal Implicit Eye Tracking Calibration for {XR}},
  author={Hou, Baosheng James and Abramyan, Lucy and Gurumurthy, Prasanthi and Adams, Haley and Tosic Rodgers, Ivana and Gonzalez, Eric J. and Patel, Khushman and Cola{\c{c}}o, Andrea and Pfeuffer, Ken and Gellersen, Hans and Ahuja, Karan and Gonzalez-Franco, Mar},
  booktitle={Proceedings of the 2025 CHI Conference on Human Factors in Computing Systems},
  articleno={550},
  pages={550:1--550:16},
  year={2025},
  doi={10.1145/3706598.3713461}
}

@inproceedings{davidjohn2021intent,
  title={Towards Gaze-Based Prediction of the Intent to Interact in Virtual Reality},
  author={David-John, Brendan and Peacock, Candace and Zhang, Ting and Murdison, T. Scott and Benko, Hrvoje and Jonker, Tanya R.},
  booktitle={ACM Symposium on Eye Tracking Research and Applications},
  articleno={2},
  pages={2:1--2:7},
  year={2021},
  doi={10.1145/3448018.3458008}
}

@inproceedings{wei2023targetselection,
  title={Predicting Gaze-Based Target Selection in Augmented Reality Headsets Based on Eye and Head Endpoint Distributions},
  author={Wei, Yushi and Shi, Rongkai and Yu, Difeng and Wang, Yihong and Li, Yue and Yu, Lingyun and Liang, Hai-Ning},
  booktitle={Proceedings of the 2023 CHI Conference on Human Factors in Computing Systems},
  articleno={283},
  pages={283:1--283:14},
  year={2023},
  doi={10.1145/3544548.3581042}
}

@inproceedings{burlingham2024dependencies,
  title={Real-World Scanpaths Exhibit Long-Term Temporal Dependencies: Considerations for Contextual {AI} for {AR} Applications},
  author={Burlingham, Charlie S. and Sendhilnathan, Naveen and Wu, Xiuyun and Murdison, T. Scott and Proulx, Michael J.},
  booktitle={Proceedings of the 2024 Symposium on Eye Tracking Research and Applications},
  articleno={89},
  pages={89:1--89:7},
  year={2024},
  doi={10.1145/3649902.3656352}
}

@article{hu2021fixationnet,
  title={FixationNet: Forecasting Eye Fixations in Task-Oriented Virtual Environments},
  author={Hu, Zhiming and Bulling, Andreas and Li, Sheng and Wang, Guoping},
  journal={IEEE Transactions on Visualization and Computer Graphics},
  volume={27},
  number={5},
  pages={2681--2690},
  year={2021},
  doi={10.1109/TVCG.2021.3067779}
}

@inproceedings{rolff2022gazetransformer,
  title={GazeTransformer: Gaze Forecasting for Virtual Reality Using Transformer Networks},
  author={Rolff, Tim and Harms, H. Matthias and Steinicke, Frank and Frintrop, Simone},
  booktitle={Pattern Recognition},
  series={Lecture Notes in Computer Science},
  volume={13485},
  pages={577--593},
  year={2022},
  publisher={Springer},
  doi={10.1007/978-3-031-16788-1_35}
}

@inproceedings{melnyk2026timeseries,
  title={Gaze Prediction as Time-Series Forecasting for Virtual Reality Applications: Quantifying Performance Variability and Extreme-Case Errors},
  author={Melnyk, Kateryna and Friedman, Lee and Komogortsev, Oleg},
  booktitle={Proceedings of the 2026 Symposium on Eye Tracking Research and Applications},
  articleno={16},
  pages={16:1--16:10},
  year={2026},
  doi={10.1145/3797246.3803043}
}

@inproceedings{kumar2008gaze,
  title={Improving the Accuracy of Gaze Input for Interaction},
  author={Kumar, Manu and Klingner, Jeff and Puranik, Rohan and Winograd, Terry and Paepcke, Andreas},
  booktitle={Proceedings of the 2008 Symposium on Eye Tracking Research \& Applications},
  pages={65--68},
  year={2008},
  doi={10.1145/1344471.1344488}
}

@article{hu2019sgaze,
  title={{SGaze}: A Data-Driven Eye-Head Coordination Model for Realtime Gaze Prediction},
  author={Hu, Zhiming and Zhang, Congyi and Li, Sheng and Wang, Guoping and Manocha, Dinesh},
  journal={IEEE Transactions on Visualization and Computer Graphics},
  volume={25},
  number={5},
  pages={2002--2010},
  year={2019},
  doi={10.1109/TVCG.2019.2899187}
}

@article{hu2020dgaze,
  title={{DGaze}: {CNN}-Based Gaze Prediction in Dynamic Scenes},
  author={Hu, Zhiming and Li, Sheng and Zhang, Congyi and Yi, Kangrui and Wang, Guoping and Manocha, Dinesh},
  journal={IEEE Transactions on Visualization and Computer Graphics},
  volume={26},
  number={5},
  pages={1902--1911},
  year={2020},
  doi={10.1109/TVCG.2020.2973473}
}

@inproceedings{casiez2012oneeuro,
  author    = {Casiez, G{\'e}ry and Roussel, Nicolas and Vogel, Daniel},
  title     = {1 Euro Filter: A Simple Speed-Based Low-Pass Filter for Noisy Input in Interactive Systems},
  booktitle = {Proceedings of the SIGCHI Conference on Human Factors in Computing Systems},
  pages     = {2527--2530},
  year      = {2012},
  publisher = {ACM},
  doi       = {10.1145/2207676.2208639}
}

@article{isomoto2022dwell,
  author  = {Isomoto, Toshiya and Yamanaka, Shota and Shizuki, Buntarou},
  title   = {Dwell Selection with ML-Based Intent Prediction Using Only Gaze Data},
  journal = {Proceedings of the ACM on Interactive, Mobile, Wearable and Ubiquitous Technologies},
  volume  = {6},
  number  = {3},
  articleno = {120},
  pages   = {120:1--120:21},
  year    = {2022},
  doi     = {10.1145/3550301}
}

@article{narkar2024gazeintent,
  author  = {Narkar, Anish S. and Michalak, Jan J. and Peacock, Candace E. and David-John, Brendan},
  title   = {GazeIntent: Adapting Dwell-Time Selection in VR Interaction with Real-Time Intent Modeling},
  journal = {Proceedings of the ACM on Human-Computer Interaction},
  volume  = {8},
  number  = {ETRA},
  articleno = {226},
  pages   = {226:1--226:18},
  year    = {2024},
  doi     = {10.1145/3655600}
}

@inproceedings{salvucci2000identifying,
  author    = {Salvucci, Dario D. and Goldberg, Joseph H.},
  title     = {Identifying Fixations and Saccades in Eye-Tracking Protocols},
  booktitle = {Proceedings of the 2000 Symposium on Eye Tracking Research \& Applications},
  pages     = {71--78},
  year      = {2000},
  publisher = {ACM},
  doi       = {10.1145/355017.355028}
}

@inproceedings{vaswani2017attention,
  author    = {Vaswani, Ashish and Shazeer, Noam and Parmar, Niki and Uszkoreit, Jakob and Jones, Llion and Gomez, Aidan N. and Kaiser, {\L}ukasz and Polosukhin, Illia},
  title     = {Attention Is All You Need},
  booktitle = {Advances in Neural Information Processing Systems},
  volume    = {30},
  year      = {2017}
}

@inproceedings{wu2023timesnet,
  author    = {Wu, Haixu and Hu, Tengge and Liu, Yong and Zhou, Hang and Wang, Jianmin and Long, Mingsheng},
  title     = {TimesNet: Temporal 2D-Variation Modeling for General Time Series Analysis},
  booktitle = {International Conference on Learning Representations},
  year      = {2023}
}

@inproceedings{lin2017focal,
  author    = {Lin, Tsung-Yi and Goyal, Priya and Girshick, Ross and He, Kaiming and Doll{\'a}r, Piotr},
  title     = {Focal Loss for Dense Object Detection},
  booktitle = {Proceedings of the IEEE International Conference on Computer Vision},
  pages     = {2980--2988},
  year      = {2017},
  doi       = {10.1109/ICCV.2017.324}
}

@inproceedings{kendall2018multitask,
  author    = {Kendall, Alex and Gal, Yarin and Cipolla, Roberto},
  title     = {Multi-Task Learning Using Uncertainty to Weigh Losses for Scene Geometry and Semantics},
  booktitle = {Proceedings of the IEEE Conference on Computer Vision and Pattern Recognition},
  pages     = {7482--7491},
  year      = {2018},
  doi       = {10.1109/CVPR.2018.00781}
}

@inproceedings{pfeuffer2017gazepinch,
author = {Pfeuffer, Ken and Mayer, Benedikt and Mardanbegi, Diako and Gellersen, Hans},
title = {Gaze + pinch interaction in virtual reality},
year = {2017},
isbn = {9781450354868},
publisher = {Association for Computing Machinery},
address = {New York, NY, USA},
url = {https://doi.org/10.1145/3131277.3132180},
doi = {10.1145/3131277.3132180},
booktitle = {Proceedings of the 5th Symposium on Spatial User Interaction},
pages = {99–108},
numpages = {10},
location = {Brighton, United Kingdom},
series = {SUI '17}
}

@inproceedings{yang2024gazetarget,
  title={Gaze target detection by merging human attention and activity cues},
  author={Yang, Yaokun and Yin, Yihan and Lu, Feng},
  booktitle={Proceedings of the AAAI Conference on Artificial Intelligence},
  volume={38},
  number={7},
  pages={6585--6593},
  year={2024}
}

@inproceedings{cai2024msgnet,
  title={Msgnet: Learning multi-scale inter-series correlations for multivariate time series forecasting},
  author={Cai, Wanlin and Liang, Yuxuan and Liu, Xianggen and Feng, Jianshuai and Wu, Yuankai},
  booktitle={Proceedings of the AAAI conference on artificial intelligence},
  volume={38},
  number={10},
  pages={11141--11149},
  year={2024}
}
\end{document}